# SOUND EVENT DETECTION VIA DILATED CONVOLUTIONAL RECURRENT NEURAL NETWORKS

*Yanxiong Li[1], Mingle Liu[1], Konstantinos Drossos[2], Tuomas Virtanen[2]*

[1]School of Electronic and Information Engineering, South China University of Technology, Guangzhou, China
[2]Audio Research Group, Tampere University, Tampere, Finland
eeyxli@scut.edu.cn, mingle96@foxmail.com, {konstantinos.drossos, tuomas.virtanen}@tuni.fi

## ABSTRACT

Convolutional recurrent neural networks (CRNNs) have achieved state-of-the-art performance for sound event detection (SED). In this paper, we propose to use a dilated CRNN, namely a CRNN with a dilated convolutional kernel, as the classifier for the task of SED. We investigate the effectiveness of dilation operations which provide a CRNN with expanded receptive fields to capture long temporal context without increasing the amount of CRNN's parameters. Compared to the classifier of the baseline CRNN, the classifier of the dilated CRNN obtains a maximum increase of 1.9%, 6.3% and 2.5% at $F$1 score and a maximum decrease of 1.7%, 4.1% and 3.9% at error rate (*ER*), on the publicly available audio corpora of the TUT-SED Synthetic 2016, the TUT Sound Event 2016 and the TUT Sound Event 2017, respectively.

## 1. INTRODUCTION

The goal of SED is to detect the activity of target sound events in audio recordings. SED can be applied to many areas related to machine listening, such as traffic monitoring [1], smart meeting room [2], automatic assistance driving [3], and multimedia analysis [4]. The popular classifiers for SED include deep models, such as CRNNs [5, 6], recurrent neural networks (RNNs) [7, 8], convolutional neural networks (CNNs) [9-10]; and traditional shallow models, such as random regression forests [11], support vector machines [12-14], hidden Markov models [15], and Gaussian mixture models [16].

The RNNs have been proven quite promising for SED [7, 8] since they are able to effectively model the temporal context of sound events. The RNN has been combined with convolutional layers, resulting in a CRNN which achieved state-of-the-art results in detecting sound events [5, 6, 17].

The dilated convolutional kernel [18] is able to increase the size of the receptive field without introducing extra parameters. Hence, to obtain the same size of receptive field, the CRNN with dilated convolutional kernels (called dilated CRNN) uses much less layers than the CRNN with conventional convolutional kernels (called baseline CRNN), which is able to avoid the overfitting problem caused by deeper networks with a lot of parameters. In addition, with the same number of parameters, the networks with the dilated convolutions are able to capture longer temporal context than that with the conventional convolutions. Modeling the long temporal context by the dilated convolutions has proved to be helpful for improving the performance of some tasks. For example, Tan et al [19] used gated residual networks with dilated convolutions to model the long temporal context, and their networks outperformed the networks without dilated convolutions for monaural speech enhancement. To the best of our knowledge, however, no effort has been made on investigating the effectiveness of dilation operations for SED.

In this paper, we propose a sound event detection system which uses dilated convolutions in order to model the long temporal context. The contributions of this study are as follows. First, we propose to use a dilated CRNN as classifier fed by the feature of log mel-band energies for SED. Second, we conduct experiments on three public audio corpora to compare the performance of dilated CRNN with that of baseline CRNN. We find that the dilated CRNNs constantly outperform the equivalent baseline CRNNs and dilation operation in convolutional layers is helpful for improving the performance of the classifier of the CRNN.

## 2. METHOD

The features used in this work are log mel-band energies which were previously adopted in [5, 17, 20], whereas dilated CRNN is proposed to be used as the classifier. CRNNs will be described in next subsections.

### 2.1. Baseline CRNN

In a baseline CRNN, the size of the receptive fields is

---
This work was partly supported by the national natural science foundation of China (61771200, 6191101570), the project of international science and technology collaboration of Guangdong (2019A050509001), and the European Unions H2020 Framework Programme through ERC Grant Agreement 637422 EVERYSOUND.

increased along layers via conventional convolutions. The baseline CRNN used in this work consists of conventional CNN blocks, bidirectional long short term memory (BLSTM) block, and sigmoid layer, as shown in Fig. 1 (a). The conventional CNN block used in this work is composed of a conventional convolutional operation, a rectification linear activation function, a pooling function, an operation of batch normalization, and a dropout operation.

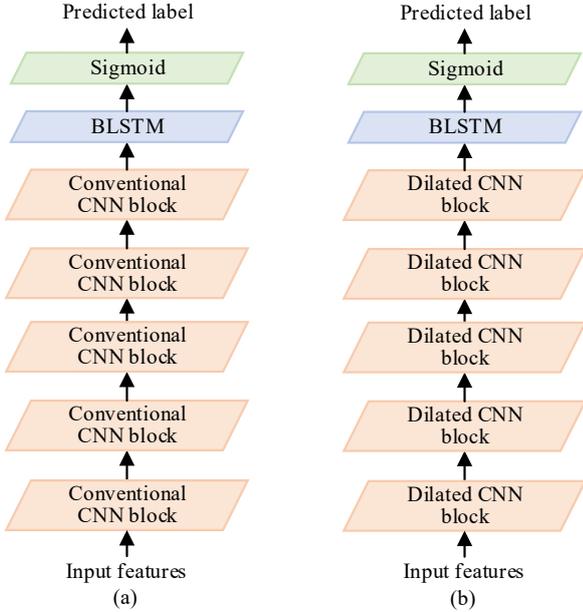

Fig. 1. The diagrams of: (a) baseline CRNN, and (b) dilated CRNN.

In the baseline CRNN, low-level and high-level conventional CNN blocks are able to extract low-level transformed features and semantic features, respectively. Then, a BLSTM block as a RNN is applied to model sequential properties of sound events in both forward and backward directions with two separate hidden layers that are connected to the same output layer [21]. Finally, a sigmoid layer is utilized to predict the probabilities of sound events.

## 2.2. Dilated CRNN

The dilated CRNN used in this work is demonstrated in Fig. 1 (b), in which the dilated CNN block consists of a dilated convolutional operation, a rectification linear activation function, a pooling function, an operation of batch normalization, and a dropout operation.

Dilated convolution was originally designed for wavelet decomposition [22]. A two-dimensional discrete convolution operator $*$ [19], which convolves the log mel-band energies $F$ with kernel $K$ of size $(2m+1) \times (2m+1)$, is defined by

$$(F * k)(p) = \sum_{s+t=p} F(s)K(t), \quad (1)$$

where $p, s \in \mathbb{Z}^2, t \in [-m, m]^2 \cap \mathbb{Z}^2$, and $\mathbb{Z}$ stands for the set of integers. The dilated version of the convolutional operator $*$, marked $*_r$, is defined by

$$(F *_r k)(p) = \sum_{s+rt=p} F(s)K(t), \quad (2)$$

where $p, s \in \mathbb{Z}^2, t \in [-m, m]^2 \cap \mathbb{Z}^2$, $r$ denotes a dilation rate, and $*_r$ is called a $r$-dilated convolution. Hence, conventional convolution can be regarded as a one-dilated convolution. Similarly, the expression of a one-dimension $r$-dilated convolution is the same to Eq. (2), except the domain of definition of the variables, i.e., $p, s \in \mathbb{Z}, t \in [-m, m] \cap \mathbb{Z}$ in the one-dimension $r$-dilated convolution. The stride of convolutions is set to 1 for keeping the time resolution the same as in the input.

Fig. 2 (a) and (b) present a conceptual illustration of conventional and dilated convolutions, respectively. The convolutions in the first layer where $r_1 = 1$, are a 3×1 convolution, and the effective receptive field of each neuron covers 3 audio frames in both Fig. 1 (a) and (b). In the second layer where $r_2 = 2$ in Fig. 2 (b) ($r_2 = 1$ in Fig. 2(a)), the effective receptive field of each neuron covers 7 audio frames in Fig. 2 (b) (5 audio frames in Fig. 2(a)). In the third (top) layer where $r_3 = 4$ in Fig. 2 (b) ($r_3 = 1$ in Fig. 2(a)), the effective receptive field of each neuron covers 15 audio frames in Fig. 2 (b) (7 audio frames in Fig. 2(a)). In summary, with the increment of the number of conventional convolutional layers (dilation rate is fixed to 1), the size of the receptive fields is expanded linearly (as shown in Fig. 2(a)). However, with the increment of the number of dilated convolutional layers, the size of the receptive fields is expanded exponentially when the kernels are applied with exponentially increasing dilation rates (as shown in Fig. 2(b)). Hence, the dilated CRNN is able to cover a wider range of information than the baseline CRNN when they have the same number of convolutional layers (i.e., the same number of parameters).

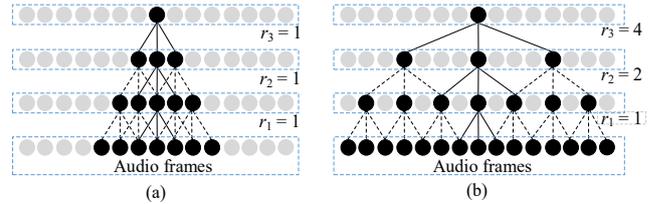

Fig. 2. Conceptual illustration of: (a) conventional convolution, and (b) dilated convolution. $r_3$, $r_2$ and $r_1$ denote dilation rates of the third (top), second and first convolutional layers, respectively.

## 3. EXPERIMENTS

The proposed method is evaluated on three publicly available audio corpora: the TUT-SED Synthetic 2016, the TUT Sound Event 2016, and the TUT Sound Event 2017 using the metrics of $F1$ score and $ER$. These three audio corpora were popularly adopted in previous works.

### 3.1. Experimental data

The TUT-SED Synthetic 2016 [5, 20] consists of 100 audio recordings which are synthetically created using 16 classes of isolated sound events: alarms/sirens, baby crying, bird singing, bus, cat meowing, crowd applause, crowd cheering, dog barking, footsteps, glass smash, gun shot, horse walk, mixer, motorcycle, rain, and thunder. Each audio recording contains a maximum of five target sound events which are randomly selected. The length of all audio recordings is 566 minutes in total. About 60% and 20% of the audio recordings are used as training data, and validation data, respectively. The rest 20% of the audio recordings are used as test data. The detailed information concerning this audio corpus can be found online[1].

Two real-life audio corpora: the TUT Sound Events 2016 and the TUT Sound Events 2017 [5, 20], were used in the experiments. These two audio corpora were extracted from the DCASE (Detection and Classification of Acoustic Scenes and Events) challenge datasets in real life audio.

The audio corpus of the TUT Sound Events 2016 contains sound events acquired in two acoustic environments: Home and Residential area. The audio recordings of Home environment contain 11 classes of sound events: rustling, snapping, cupboard, cutlery, dishes, drawer, glass jingling, object impact, people walking, washing dishes, and water tap running. The audio recordings of Residential area environment have 7 classes of sound events: banging, bird singing, car passing by, children shouting, people speaking, people walking, and wind blowing.

The audio corpus of the TUT Sound Events 2017 contains audio recordings acquired in a street environment. This audio corpus contains 6 different classes of sound events: brakes squeaking, car, children, large vehicle, people speaking, and people walking.

We use the cross-fold validation division adopted in the DCASE 2016 and 2017 challenges for the audio corpora of both the TUT Sound Events 2016 and the TUT Sound Events 2017. The detailed information concerning the classes of sound events, the settings of cross-fold division, and the audio recordings can be found online[2, 3].

### 3.2. Parameters settings and metrics

Each audio recording is split into audio frames of 20 ms with frame overlapping of 50%, for extracting the feature of log mel-band energies with 40 dimensions. The values of dilation rate range from 2 to 32. The size of convolutional kernels is: 3 × 3. The state number of BLSTM module and dropout rate are experimentally set to 128 and 10%, respectively. The value of learning rate is initially set to 0.01 and then dynamically attenuated. The rectification linear activation function is the ReLU. To optimize the weights of neural networks, the Adam optimizer [23] is employed with default values. The size of output layer of CRNNs is determined by the class number of sound events in each audio corpus, such as 16 for the TUT-SED Synthetic 2016. A batch size of 16 is used and the training procedure is stopped when the loss for the validation data is not decreasing for 30 consecutive epochs.

The performance of the methods is evaluated using the frame based $F1$ score and $ER$, which were popularly adopted in previous works [20]. The detailed information for calculating $F1$ score and $ER$ is referred to [5].

### 3.3. Results and discussions

We discuss the performance of individual CRNNs with different settings of convolutional layers for SED. The difference between baseline CRNN and dilated CRNN consists in the dilation rates only. That is, in each convolutional layer, the dilation rate is fixed to 1 for baseline CRNN and is variable for dilated CRNN. The dilation operation is conducted in the directions of both time and frequency. The $F1$ scores and $ER$s obtained by the CRNNs with different settings of convolutional layers are given in Tables 1 to 3 evaluated on the Synthetic 2016, the TUT Sound Events 2016, and the TUT Sound Events 2017, respectively. In the column of "Dilation rate", the digits denote dilation rates with the increase of number of convolutional layers. For example, 2-4-8-16-32 stands for there are 5 convolutional layers in the CRNN and the dilation rates of the first, second, third, fourth and fifth convolutional layers are 2, 4, 8, 16 and 32, respectively.

First, we evaluate the methods for SED on the audio corpus of the Synthetic 2016. It can be seen from Table 1 that the dilated CRNN achieves the maximum $F1$ score of 59.4% (higher is better), and the minimum $ER$ of 48.6% (lower is better) when the dilation rate is set to 2-4-8. In addition, the dilated CRNNs always outperform the equivalent baseline CRNNs (with the same number of convolutional layers) in terms of both $F1$ score and $ER$. Compared to the baseline CRNN, the dilated CRNN obtains a maximum increase of 1.9% at $F1$ score and a maximum decrease of 1.7% at $ER$.

Table 1. $F1$ scores and $ER$s obtained by baseline and dilated CRNNs with various dilation rates on the Synthetic 2016 (in %).

| CRNNs | | Synthetic 2016 | |
|---|---|---|---|
| Network no. | Dilation rate | F1 | ER |
| Baseline CRNN1 | 1 | 56.8 | 50.8 |
| Dilated CRNN1 | 2 | 58.6 | 49.9 |
| Baseline CRNN2 | 1-1 | 57.0 | 51.3 |
| Dilated CRNN2 | 2-4 | 58.9 | 49.6 |
| Baseline CRNN3 | 1-1-1 | 58.2 | 49.5 |
| Dilated CRNN3 | 2-4-8 | **59.4** | **48.6** |
| Baseline CRNN4 | 1-1-1-1 | 57.5 | 50.5 |
| Dilated CRNN4 | 2-4-8-16 | 57.8 | 50.4 |
| Baseline CRNN5 | 1-1-1-1-1 | 57.7 | 49.9 |
| Dilated CRNN5 | 2-4-8-16-32 | 58.5 | 49.8 |

Then, we evaluate the methods on the audio corpus of the TUT Sound Events 2016. There are two subsets in this audio corpus, i.e., Home and Residential area. The results in Table 2 show that the dilated CRNN obtains the maximum $F$1 scores of 42.8% and 65.4%, and the minimum $ER$s of 64.6% and 42.0%, when the dilation rate is set to 2-4-8 and 2-4 for the subsets of Home and Residential area, respectively. Compared to the baseline CRNN, the dilated CRNN obtains a maximum increase of 6.3% and 2.0% at $F$1 score and a maximum decrease of 4.1% and 2.4% at $ER$, for the subsets of Home and Residential area, respectively. In addition, the dilated CRNNs constantly have better performance compared to the equivalent baseline CRNNs.

---
[1] http://www.cs.tut.fi/sgn/arg/taslp2017-crnn-sed/tut-sed-synthetic-2016
[2] http://www.cs.tut.fi/sgn/arg/dcase2016
[3] http://www.cs.tut.fi/sgn/arg/dcase2017/challenge

Table 2. $F$1 scores and $ER$s obtained by baseline and dilated CRNNs with various dilation rates on the TUT Sound Events 2016 (in %).

| CRNNs | | TUT Sound Events 2016 | | | |
|---|---|---|---|---|---|
| | | Home | | Residential area | |
| Network no. | Dilation rate | $F$1 | $ER$ | $F$1 | $ER$ |
| Baseline CRNN1 | 1 | 35.5 | 70.0 | 62.8 | 45.1 |
| Dilated CRNN1 | 2 | 41.8 | 65.9 | 63.4 | 44.1 |
| Baseline CRNN2 | 1-1 | 37.7 | 70.4 | 63.4 | 43.9 |
| Dilated CRNN2 | 2-4 | 40.0 | 67.5 | **65.4** | **42.0** |
| Baseline CRNN3 | 1-1-1 | 41.1 | 66.6 | 62.4 | 45.3 |
| Dilated CRNN3 | 2-4-8 | **42.8** | **64.6** | 64.3 | 42.9 |
| Baseline CRNN4 | 1-1-1-1 | 37.0 | 68.9 | 61.2 | 46.3 |
| Dilated CRNN4 | 2-4-8-16 | 39.0 | 67.5 | 62.4 | 45.3 |
| Baseline CRNN5 | 1-1-1-1-1 | 37.9 | 68.3 | 62.3 | 45.2 |
| Dilated CRNN5 | 2-4-8-16-32 | 42.2 | 66.2 | 63.9 | 43.6 |

Finally, the results obtained by the methods for SED on the audio corpus of the TUT Sound Events 2017 are presented in Table 3. The dilated CRNN achieves the maximum $F$1 score of 60.7%, and the minimum $ER$ of 49.5% when the dilation rate is set to 2-4-8-16-32. Compared to the baseline CRNN, the dilated CRNN obtains a maximum increase of 2.5% at $F$1 score and a maximum decrease of 3.9% at $ER$. In addition, the dilated CRNNs are always superior to the equivalent baseline CRNNs in terms of $F$1 score and $ER$.

Table 3. $F$1 scores and $ER$s obtained by baseline and dilated CRNNs with various dilation rates on the TUT Sound Events 2017 (in %).

| CRNNs | | TUT Sound Events 2017 | |
|---|---|---|---|
| Network no. | Dilation rate | $F$1 | $ER$ |
| Baseline CRNN1 | 1 | 59.0 | 51.6 |
| Dilated CRNN1 | 2 | 59.3 | 49.9 |
| Baseline CRNN2 | 1-1 | 58.6 | 52.1 |
| Dilated CRNN2 | 2-4 | 60.2 | 51.0 |
| Baseline CRNN3 | 1-1-1 | 59.4 | 51.7 |
| Dilated CRNN3 | 2-4-8 | 60.4 | 50.9 |
| Baseline CRNN4 | 1-1-1-1 | 58.5 | 52.8 |
| Dilated CRNN4 | 2-4-8-16 | 60.6 | 50.5 |
| Baseline CRNN5 | 1-1-1-1-1 | 58.2 | 53.4 |
| Dilated CRNN5 | 2-4-8-16-32 | **60.7** | **49.5** |

In summary, the dilated CRNNs obtain higher $F$1 scores and lower $ER$s than the equivalent baseline CRNNs on three audio corpora. With different settings of dilation rates, the dilated CRNN has larger receptive fields for capturing long temporal context which is beneficial for improving the performance of the CRNNs for SED.

The results presented in Tables 1 to 3 are obtained by the CRNNs which are the final models after convergence during training procedure. In order to show the performance of the baseline and dilated CRNNs obtained at different training epochs, we depict $F$1 scores and $ER$s achieved by some representative CRNNs (evaluated on the test data), namely, baseline CRNN3 and dilated CRNN3 in Table 1, baseline CRNN3 and dilated CRNN3 (for the Home) in Table 2, baseline CRNN2 and dilated CRNN2 (for the Residential area) in Table 2, baseline CRNN5 and dilated CRNN5 in Table 3. It can be seen from Fig. 3 that the dilated CRNNs obtain higher $F$1 scores and lower $ER$s than the equivalents baseline CRNNs in most epochs. That is, the dilated CRNNs obtained at different training epochs outperforms the equivalent baseline CRNNs in most time when they are evaluated on the test data.

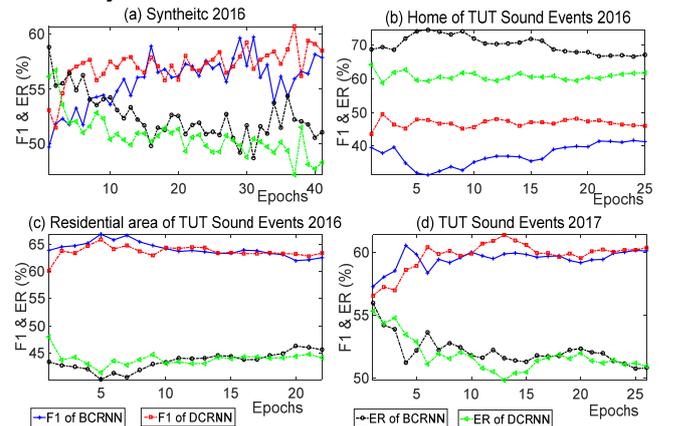

Fig. 3. The $F$1 scores and $ER$s obtained by the baseline and dilated: (a) CRNN3 in Table 1 on the Synthetic 2016, (b) CRNN3 in Table 2 on the Home of the TUT Sound Events 2016, (c) CRNN2 in Table 2 on the Residential area of the TUT Sound Events 2016, and (d) CRNN5 in Table 3 on the TUT Sound Events 2017, at different training epochs. BCRNN: baseline CRNN; DCRNN: dilated CRNN.

## 4. CONCLUSIONS

In this work, we conducted a preliminary study for SED via investigating the effectiveness of dilation operations in convolutional layers of CRNNs. Evaluated on three public audio corpora, the experimental results showed that the dilated CRNNs always outperforms their equivalent baseline CRNNs in terms of both $F$1 score and $ER$s. That is, dilation operation with different rates has contribution to improve the performance of the CRNNs for the task of SED.

In this study, we focused only on the effectiveness of dilation rates. In future work, we will discuss the influence of other parameters of CRNN on the performance of SED, such as pooling, activation function, dropout, initial weights.